\documentclass[]{ws-ijbc1}

\usepackage{url}
\usepackage{subfigure}

\pdfoutput=1

\begin{document}

%\begin{frontmatter}

%\catchline{}{}{}{}{} % Publisher's Area please ignore

\markboth{Adamatzky, De Lacy Costello, Bull}{On polymorphic gates in sub-excitable chemical medium}

\title{ON POLYMORPHIC LOGICAL GATES IN SUB-EXCITABLE CHEMICAL MEDIUM}

\author{Andrew Adamatzky, Ben De Lacy Costello, Larry Bull}

\address{University of the West of England, Bristol, United Kingdom}

%\date{\today}

\maketitle

%\begin{history}
%\received{(to be inserted by publisher)}
%\end{history}

\begin{abstract}

\noindent
In a sub-excitable light-sensitive Belousov-Zhabotinsky (BZ) chemical medium an asymmetric disturbance causes the formation
of localized traveling wave-fragments. Under the right conditions these wave-fragment can conserve their shape and 
velocity vectors for extended time periods. The size and life span of a fragment depend on the illumination level of the medium. 
When two or more wave-fragments collide they annihilate or merge 
into a new wave-fragment. In computer simulations based on the Oregonator model we demonstrate that the outcomes of 
inter-fragment collisions can be controlled by varying the illumination level applied to the medium.
We interpret these wave-fragments as values of Boolean variables and design collision-based
polymorphic logical gates. The gate implements operation {\sc xnor} for low illumination, 
and it acts as {\sc nor} gate for high illumination. As a {\sc nor} gate is a universal gate then we are able to demonstrate that a simulated light sensitive BZ medium exhibits computational universality.        

\vspace{0.5cm}

\noindent

\vspace{0.5cm}

\end{abstract}

%\pacs{82.40.-g; 82.40.Ck; 89.75.Kd; 89.75.Fb; 89.20.Ff}

%\end{frontmatter}

\section{Introduction}

Designing of logical gates in chemical systems can be traced back to early 1990s. 
Hjemfelft and colleagues suggested a theoretical coupled mass flow system 
for implementing logic gates and finite-state machines\cite{hjelmfelt_1991}, 
\cite{hjelmfelt_1992}, 
\cite{hjelmfelt_ross_1993}, \cite{hjelmfelt_1993},  \cite{hjelmfelt_ross_1995}
and Lebender and Schneider described approaches towards building logical 
gates using a series of flow rate coupled continuous stirred tank reactors and a bistable chemical
reaction\cite{lebender_1994}. No experimental prototypes were implemented at that time. 

In 1994 T\'{o}th, Showalter and Steinbock presented the first ever experimental implementation of logical gates in the 
Belousov-Zhabotinsky system\cite{toth1994}, \cite{toth_showalter}. Their constructs of logical gates were based
on the configuration of excitation wave propagation channels and the ratio between channel diameter and the critical nucleation 
radii of the excitable media. Their findings aroused great interest and resulted in several innovative designs of
computational devices, including logical gates for Boolean and 
multiple-valued logic\cite{sielewiesiuk_2001},  \cite{motoike_2003}, \cite{gorecki_2009}, \cite{yoshikawa_2009}, many-input logical gates\cite{gorecki_2006}, \cite{gorecki_2006a}, counters\cite{gorecki_2003}, coincidence detector\cite{gorecka_2003}, 
detectors of direction and distance\cite{gorecki_2005}, \cite{yoshikawa_2009a} and inductive memory\cite{motoike_2003}.
All these chemical computing devices were realised in geometrically-constrained media, where excitation waves propagated
along defined channels loaded with catalyst or tubes filled with BZ reagents. The waves perform computation by interacting 
at the junctions between the channels/tubes. Such an approach is noble, however, it essentially just imitates
conventional computing architectures (wires and valves) but using novel materials (excitable chemical systems). 
Computing in unconstrained media would be a step forward towards the implementation of massively-parallel chemical processors.

To appreciate the massive-parallelism of a thin-layer chemical media we can adopt the paradigm of collision-based computing~  \cite{cbc}. This paradigm originates from the computational universality of the Game of Life~  \cite{berlekamp_1992}, conservative
logic and the billiard-ball model  \cite{fredkin_toffoli_1982} and their cellular-automaton implementations~  \cite{margolus}. 
A collision-based computer employs mobile self-localized excitation to represent quanta of information in active non-linear media. Information values, e.g. truth values of logical variables, are given
by either the absence or presence of the localizations or by other parameters such as direction or velocity.  The localizations travel 
in space and collide with each other. The results of the collisions are interpreted as computation. There are no predetermined stationary wires, a trajectory of the travelling localization is a momentary wire. Almost any part of the reactor space can be used as a wire. Localizations can collide anywhere within this space. The localizations undergo transformations, form bound
states, annihilate or fuse when they interact. Information values of localizations are 
transformed as a result of these collisions\cite{cbc}.

To implement a collision-based scheme in a spatially-extended chemical medium we must employ travelling localisations. 
The self-localized excitation wave-fragments,  traveling in a light-sensitive BZ medium when it is in a sub-excitable state~  \cite{sendina_2001} are ideal candidates. These excitation wave-fragments behave like quasi-particles. They  exhibit rich dynamics of collisions, including quasi-reflection, fission, fusion, and annihilation\cite{andy_ben_BZ_collision}, \cite{RITABEN}. 
Using the wave-fragments we have implemented collision-based computing schemes~  \cite{adamatzky_2004_collision}. We have produced 
a range of basic collision-based computing schemes in computer simulations and chemical laboratory experiments including
logical gates\cite{adamatzky_2004_collision}, \cite{andy_ben_BZ_collision},  \cite{RITABEN}, \cite{ben_gun}, evolvable chemical logical circuits\cite{toth_2009}, and elements of a one-bit adder\cite{adamatzky_physarumgates}, 
\cite{adamatzky_adder}.

All these excitable chemical computing devices are light-sensitive, the wave-fragments grow in size with a decrease in illumination, and the wave-fragments collapse with an increase in illumination. In a very narrow illumination the wave-fragments remain localized and conserve their shape and velocity vectors for a (relatively) significant amount of time. Thus the wave-fragments can be used to represent quanta of information. What happens if the level of illumination is altered between the lower and upper limits of the wave-fragments-stability range? In the present paper we demonstrate that a design can be implemented where
the outcomes of collisions between wave-fragments sensitively depend on the level of illumination. A logical function is realised by the collision of wave fragments. By changing the illumination level we are able to change the logical function. 
Thus we implement BZ collision-based polymorphic logical gates\cite{stoika_2002}, i.e. gates which change their function depending on control signals.

The paper is structured as follows. In Sect.~\ref{methods} we show how we simulate the 
light-sensitive Belousov-Zhabotinsky system. Collisions between wave-fragments are 
studied in Sect.~\ref{collisions}. We describe our implementaton of 
collision-based polymorphic logical gates in Sect.~\ref{sectiongates}.

\section{Methods}
\label{methods}

We use the two-variable Oregonator equation\cite{field_noyes_1974} adapted to a light-sensitive 
Belousov-Zhabotinsky (BZ) reaction with applied illumination\cite{beato_engel}:

\begin{eqnarray}
  \frac{\partial u}{\partial t} & = & \frac{1}{\epsilon} (u - u^2 - (f v + \phi)\frac{u-q}{u+q}) + D_u \nabla^2 u \nonumber \\
  \frac{\partial v}{\partial t} & = & u - v 
\label{equ:oregonator}
\end{eqnarray}

The variables $u$ and $v$ represent the local concentrations of activator, or excitatory component, and inhibitor,
or refractory component. Parameter $\epsilon$ sets up a ratio of time scale for the variables $u$ and $v$, $q$ is a 
scaling parameter dependent on the rates of activation/propagation and inhibition, $f$ is a stoichiometric factor. 
Constant $\phi$ is the rate of inhibitor production. In the light-sensitive BZ $\phi$ represents the rate of inhibitor
production which is proportional to the intensity of illumination. We integrate the system using the Euler method with five-node Laplace operator, time step $\Delta t=0.005$ and grid point spacing $\Delta x= 0.25$, $\epsilon=0.022$, $f=1.4$, $q=0.002$. 
The equations effectively map the space-time dynamics of excitation in the BZ medium and have proved to be an invaluable tool 
for studying the dynamics of collisions between travelling localized excitations in our previous work~  \cite{adamatzky_2004_collision}, \cite{andy_ben_BZ_collision},  \cite{RITABEN,ben_gun}.

\begin{figure}[!tbp]
\centering
\subfigure[$\phi=0.07$]{\includegraphics[width=0.25\textwidth]{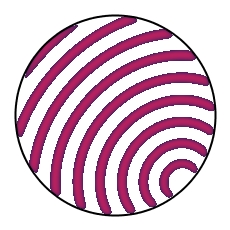}}
\subfigure[$\phi=0.077$]{\includegraphics[width=0.25\textwidth]{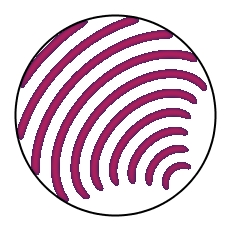}}
\subfigure[$\phi=0.07873$]{\includegraphics[width=0.25\textwidth]{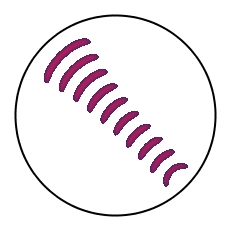}}
\subfigure[$\phi=0.07877$]{\includegraphics[width=0.25\textwidth]{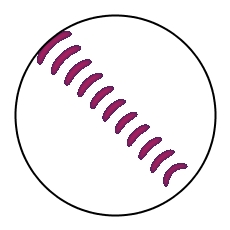}}
\subfigure[$\phi=0.07878$]{\includegraphics[width=0.25\textwidth]{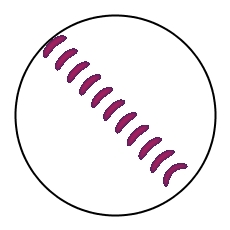}}
\subfigure[$\phi=0.079$]{\includegraphics[width=0.25\textwidth]{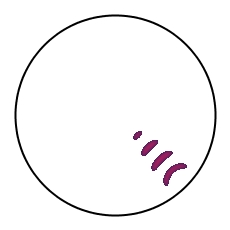}}
\caption{Showing that the development of the initial excitation is sensitively dependent on the level of illumination $\phi$. Excitation is initiated at the south-west edge of the vesicle. }
\label{singlewave}
\end{figure}

The parameter $\phi$ characterizes the excitability of the simulated medium. The medium is excitable and exhibits 
`classical' target waves, e.g. when $\phi=0.07$ (Fig.~\ref{singlewave}a) and the medium is sub-excitable with 
propagating localizations, or wave-fragments, when $\phi$ is between $0.07873$ and $0.07878$ (Fig.~\ref{singlewave}c--e).
The medium becomes non-excitable for $\phi \geq 0.79$, and after this point wave-fragments collapse after relatively short time scales (Fig.~\ref{singlewave}f).

When the BZ reaction is in a sub-excitable mode asymmetric perturbations lead to the formation of propagating localized excitation, 
or excitation wave-fragments. Wave-fragments of this type may travel in a predetermined direction for a finite period of time.
If wave-fragments kept their shape indefinitely, we would be able to build a collision-based computing circuit 
of any size.  In reality, the wave-fragments are inherently unstable: after some period of conserved-shape/ distance travelled a
wave-fragment either collapses or expands. 

Recently \cite{neuneu}, \cite{adamatzky_adder} we found a way to overcome the problem of wave-fragment instability via the 
subdivision of the computing substrate into interconnected compartments, so called BZ-vesicles,
and allowing waves to collide only inside the compartments. Each BZ-vesicle has a membrane that is impassable for excitation~  \cite{gorecki_private}, \cite{neuneu}. A pore, or a channel, between two vesicles is formed when
two vesicles come into direct contact.  The pore is small such that when a wave passes through the pore there 
is insufficient time for the wave to expand or collapse before interacting with other waves entering through adjacent pores, 
or sites of contact.

A spherical compartment --- BZ-vesicle --- is the best natural choice as it allows for effortless arrangement of the vesicles into a regular lattice, has an almost unlimited number of input/output states and also loosely conforms to a structure likely to be achieved in experiments involving the encapsulation of excitable chemical media in a lipid membrane~  \cite{gorecki_private}, \cite{neuneu}. We simulate a vesicle filled with BZ solution as a disc with radius $R$ centered in $(x_0, y_0)$. Sites inside the disc are excitable, sites outside the disc are not excitable. We imitate wave-fragment entering the vesicle by exciting (assigning values $u=1.$) grid nodes inside the small disc with radius $r$, centered in $(x_0+(R-s) cos(\theta), y_0+(R-s) sin(\theta))$. The following parameters are used in the illustrations: $R=100$, $r=5$, $s=5$, $\theta \in [0, 2\pi]$. Time lapse snapshots provided in the paper were recorded at every 150 time steps, and grid sites with excitation level $u >0.04$ were displayed.

\section{Binary collisions}
\label{collisions}

\begin{figure}[!tbp]
\centering
\subfigure[scheme]{\includegraphics[width=0.25\textwidth]{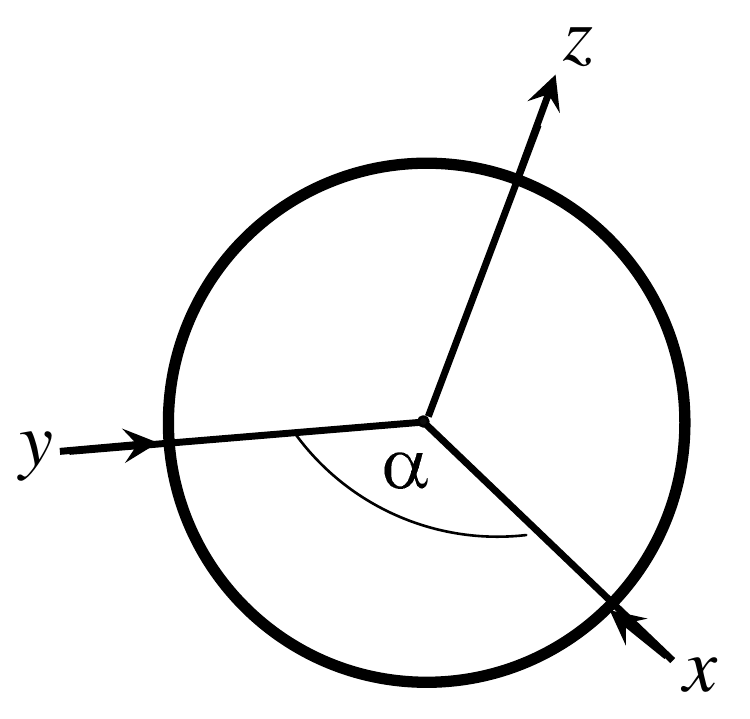}}
\subfigure[$\alpha=\frac{\pi}{18}$]{\includegraphics[width=0.19\textwidth]{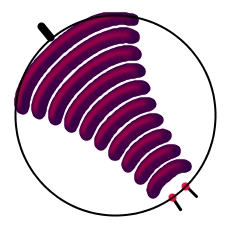}}
\subfigure[$\alpha=\frac{\pi}{9}$]{\includegraphics[width=0.19\textwidth]{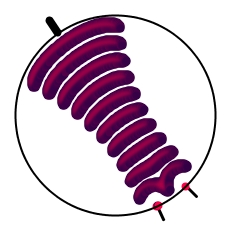}}
\subfigure[$\alpha=\frac{\pi}{6}$]{\includegraphics[width=0.19\textwidth]{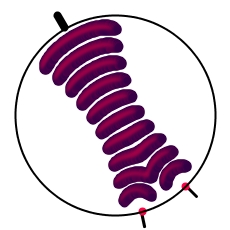}}
\subfigure[$\alpha=\frac{2\pi}{9}$]{\includegraphics[width=0.19\textwidth]{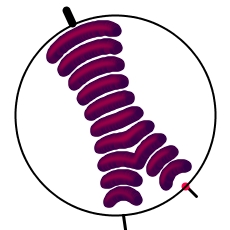}}
\subfigure[$\alpha=\frac{5\pi}{18}$]{\includegraphics[width=0.19\textwidth]{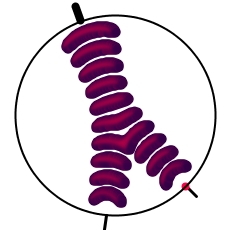}}
\subfigure[$\alpha=\frac{\pi}{3}$]{\includegraphics[width=0.19\textwidth]{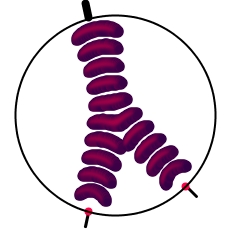}}
\subfigure[$\alpha=\frac{7\pi}{18}$]{\includegraphics[width=0.19\textwidth]{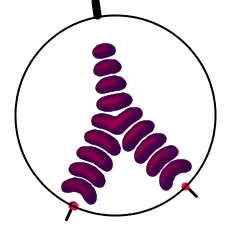}}
\subfigure[$\alpha=\frac{4\pi}{9}$]{\includegraphics[width=0.19\textwidth]{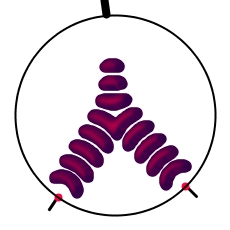}}
\subfigure[$\alpha=\frac{\pi}{2}$]{\includegraphics[width=0.19\textwidth]{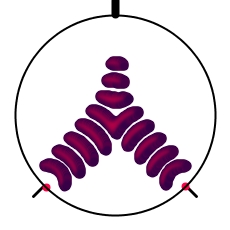}}
\subfigure[$\alpha=\frac{5\pi}{9}$]{\includegraphics[width=0.19\textwidth]{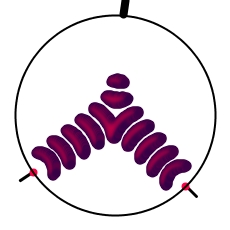}}
\subfigure[$\alpha=\frac{11\pi}{18}$]{\includegraphics[width=0.19\textwidth]{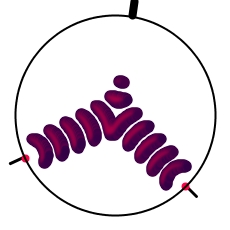}}
\subfigure[$\alpha=\frac{2\pi}{3}$]{\includegraphics[width=0.19\textwidth]{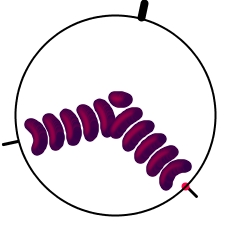}}
\subfigure[$\alpha=\frac{13\pi}{18}$]{\includegraphics[width=0.19\textwidth]{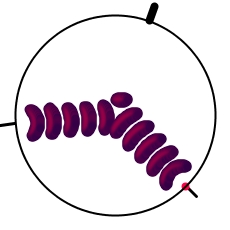}}
\subfigure[$\alpha=\frac{7\pi}{9}$]{\includegraphics[width=0.19\textwidth]{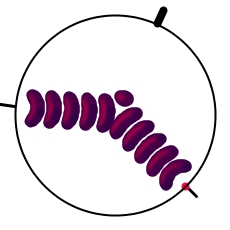}}
\subfigure[$\alpha=\frac{5\pi}{6}$]{\includegraphics[width=0.19\textwidth]{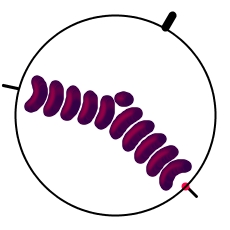}}
\subfigure[$\alpha=\frac{8\pi}{9}$]{\includegraphics[width=0.19\textwidth]{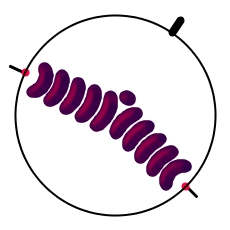}}
\subfigure[$\alpha=\frac{17\pi}{18}$]{\includegraphics[width=0.19\textwidth]{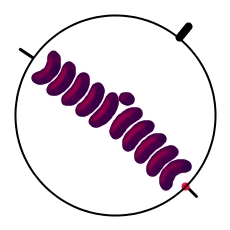}}
\subfigure[$\alpha=\pi$]{\includegraphics[width=0.19\textwidth]{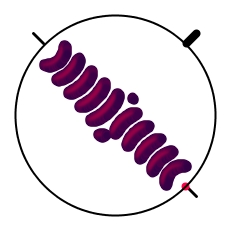}}
\caption{Outcomes of  collisions between two wave-fragments $x$ and $y$ approaching each other at angle 
$\alpha$, illumination level is $\phi=0.07873$; $z$ shows orientation of velocity vector of 
the resultant wave-fragment, produced in the collision. }
\label{catalogueofcollisions}
\end{figure}

If two wave-fragments $x$ and $y$ are initiated at the disc's edge at the same time they collide with 
each other, while approaching the centre of the disc. The outcome of the collision depends on the angle $\alpha$ between the 
velocity vectors $\vec{x}$ and $\vec{y}$  (Fig.~\ref{catalogueofcollisions}). When the angle is less than 
some critical value $\beta$ the colliding wave-fragments merge into a new wave-fragment $z$ (Fig.~\ref{catalogueofcollisions}b--g) whose velocity vector is positioned exactly between the velocity vectors of wave-fragments $x$ and $y$: $\vec{z}=(\vec{x}+\vec{y})/2$ (Fig.~\ref{catalogueofcollisions}a). When the angle  between the vectors of colliding wave-fragments exceeds some 
critical value $\beta$, the colliding wave fragments annihilate (Fig.~\ref{catalogueofcollisions}h--s). 

\begin{figure}[!tbp]
\centering
\subfigure[]{\includegraphics[width=0.7\textwidth]{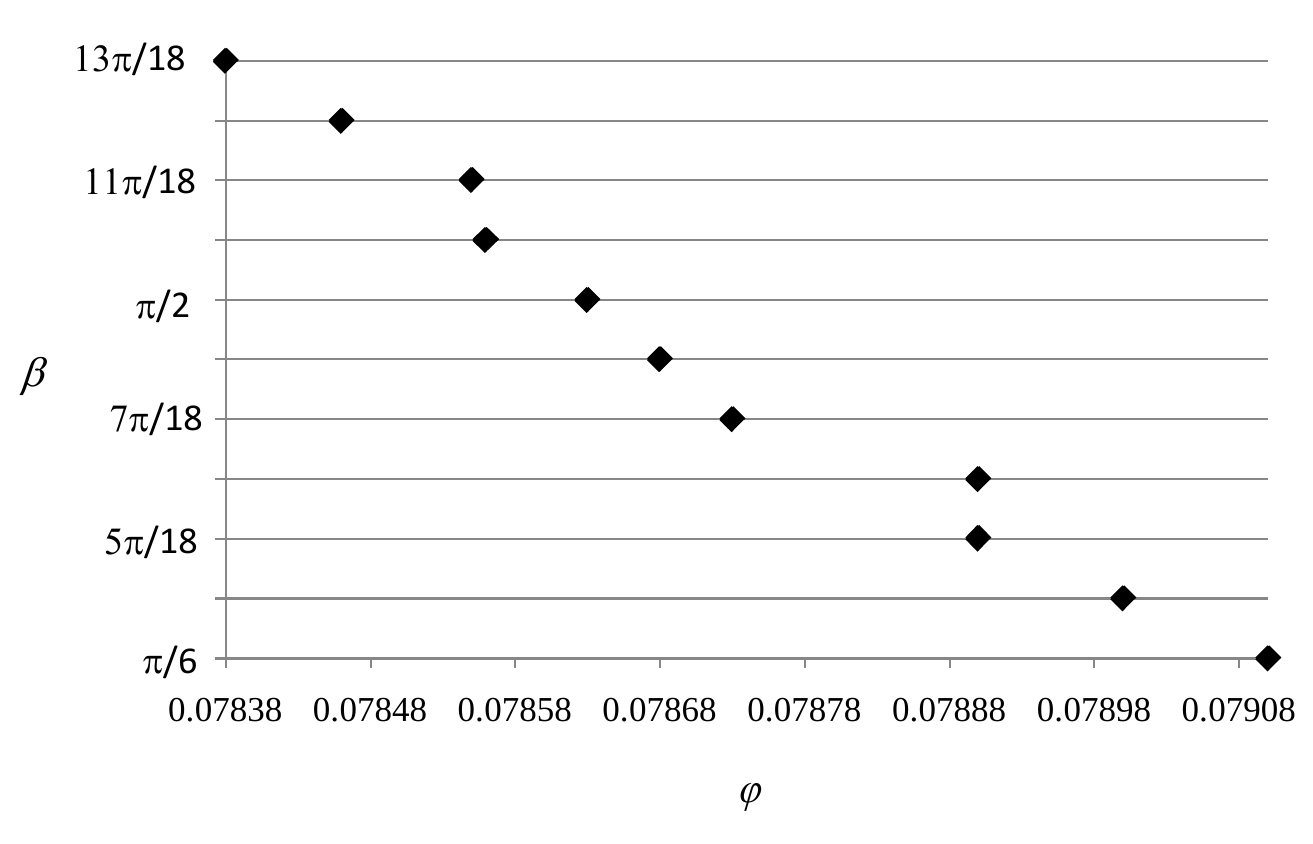}}\\
\subfigure[$\alpha=\frac{\pi}{9}$]{\includegraphics[width=0.19\textwidth]{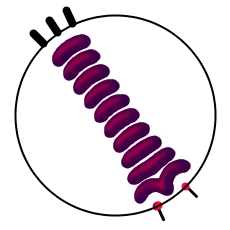}} 
%\subfigure[$\alpha=30$]{\includegraphics[width=0.25\textwidth]{figs/examplesofcriticalangles/lapse30}}
\subfigure[$\alpha=\frac{5\pi}{18}$]{\includegraphics[width=0.19\textwidth]{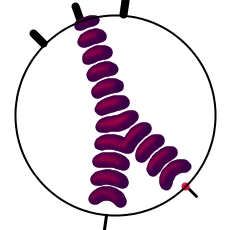}}
%\subfigure[$\alpha=60$]{\includegraphics[width=0.25\textwidth]{figs/examplesofcriticalangles/lapse60}}
%\subfigure[$\alpha=70$]{\includegraphics[width=0.25\textwidth]{figs/examplesofcriticalangles/lapse70}}
\subfigure[$\alpha=\frac{4\pi}{9}$]{\includegraphics[width=0.19\textwidth]{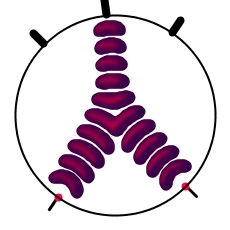}}
%\subfigure[$\alpha=90$]{\includegraphics[width=0.25\textwidth]{figs/examplesofcriticalangles/lapse90}}
%\subfigure[$\alpha=100$]{\includegraphics[width=0.25\textwidth]{figs/examplesofcriticalangles/lapse100}}
\subfigure[$\alpha=\frac{11\pi}{18}$]{\includegraphics[width=0.19\textwidth]{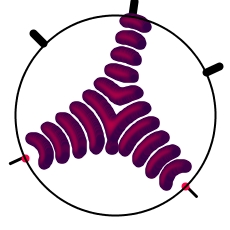}}
%\subfigure[$\alpha=140$]{\includegraphics[width=0.25\textwidth]{figs/examplesofcriticalangles/lapse140}}
\subfigure[$\alpha=\frac{5\pi}{6}$]{\includegraphics[width=0.19\textwidth]{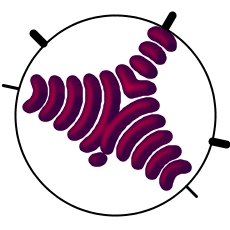}}
\caption{Level of illumination determines critical angle of wave-fragments collision:(a)~dependence of critical collision angle $\beta$ on illumination parameters $\phi$, time lapse snapshots
of wave-fragments colliding at angles $\beta$ for their critical values of illumination $\phi$:
(b)~$\alpha=\frac{\pi}{9}$, $\phi=0.0791$,
(c)~$\alpha=\frac{5\pi}{18}$, $\phi=0.07883$,
(d)~$\alpha=\frac{4\pi}{9}$, $\phi=0.07867$,
(e)~$\alpha=\frac{11\pi}{18}$, $\phi=0.07850$,
(f)~$\alpha=\frac{5\pi}{6}$, $\phi=0.078$. 
}
\label{dependencies}
\end{figure}

Let $\beta$ be a critical value such that wave-fragments colliding at angle $\alpha \leq \beta$ merge into a
wave-fragment, which propagates over an indefinitely long distance, and wave-fragments colliding at angle $\alpha > \beta$ 
annihilate.

\vspace{0.5cm}

\begin{proposition}
Critical value $\beta$ is inversely proportional to illumination level $\phi$. 
\end{proposition}

Dependence of $\beta$ on $\phi$ calculated in computational experiments is shown in Fig.~\ref{dependencies}a. The dependence 
is essentially linear, the deviations shown are due to digitization of the space in numerical experiments.
Each critical value $\beta$ has its own illumination level $\phi$, see examples in Fig.~\ref{dependencies}b--f.Therefore, by varying the illumination of the disc we can alter the outcomes of collisions between wave fragments.

\section{Polymorphic gates}
\label{sectiongates}

\begin{figure}[!tbp]
\centering
\subfigure[]{
$\stackrel{\includegraphics[scale=0.5]{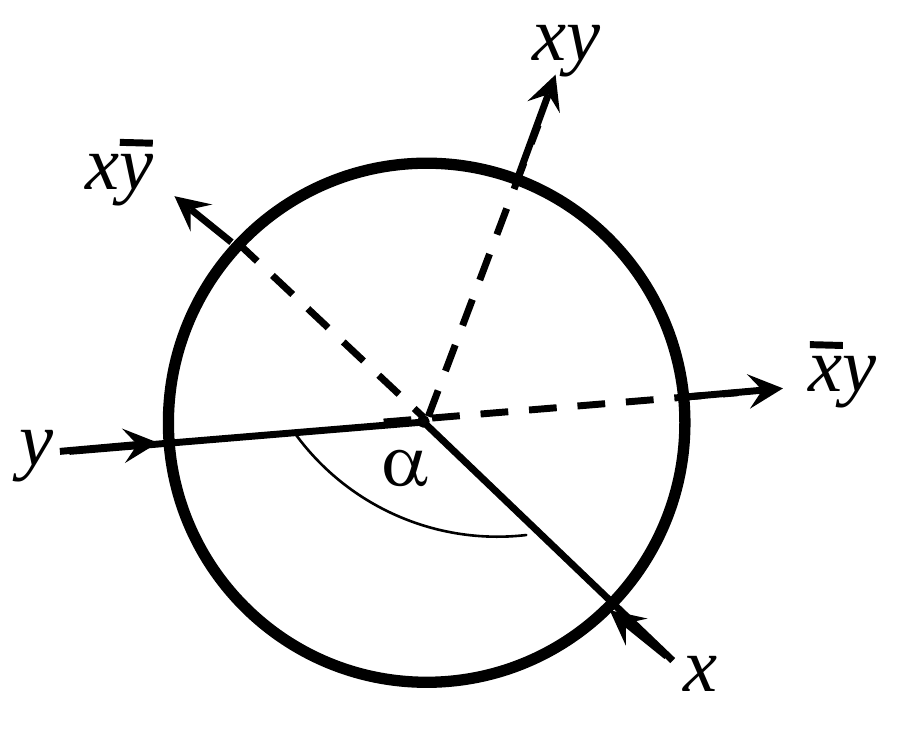}}{\text{\footnotesize low }\phi}$
$\stackrel{\includegraphics[scale=0.5]{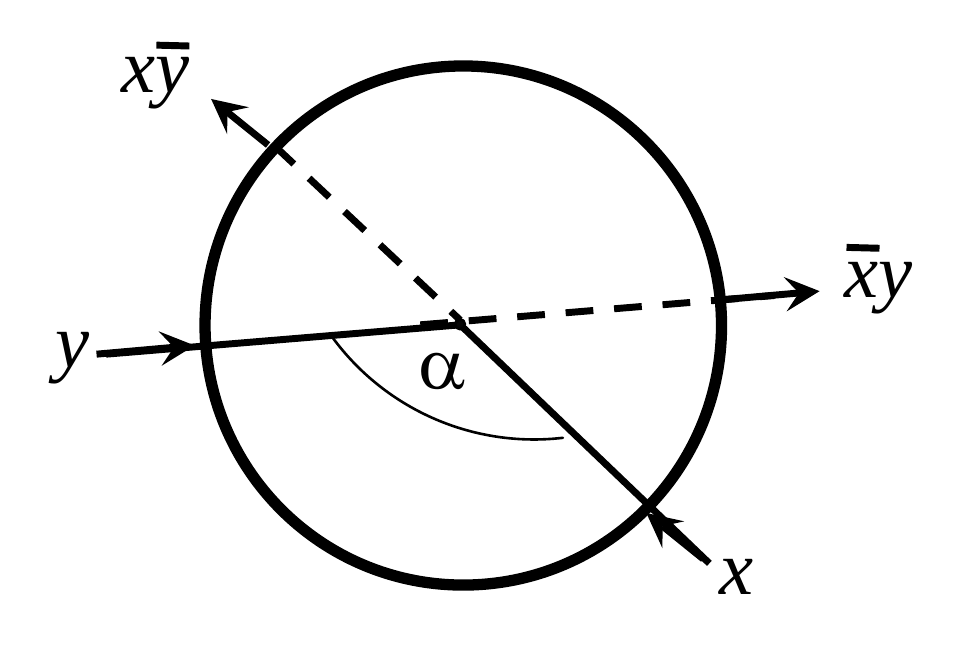}}{\text{\footnotesize high }\phi}$
}
\subfigure[]{
$\stackrel{\includegraphics[scale=0.5]{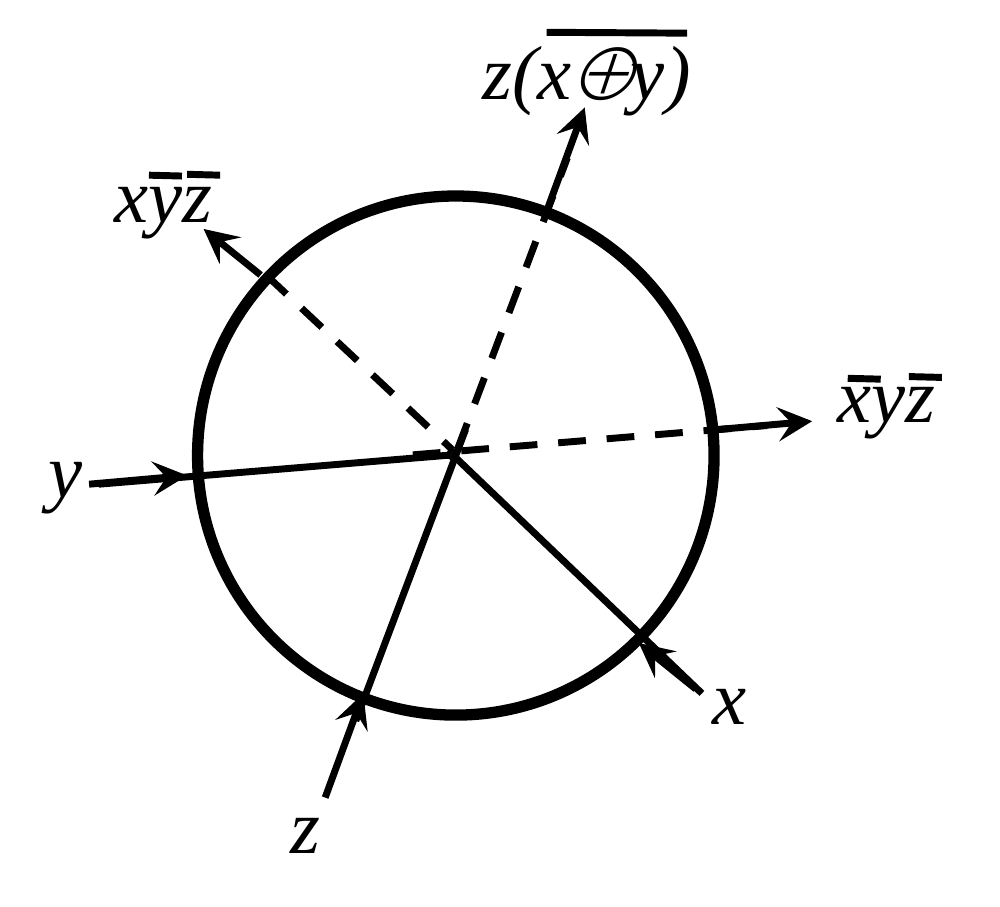}}{\text{\footnotesize low }\phi}$
$\stackrel{\includegraphics[scale=0.5]{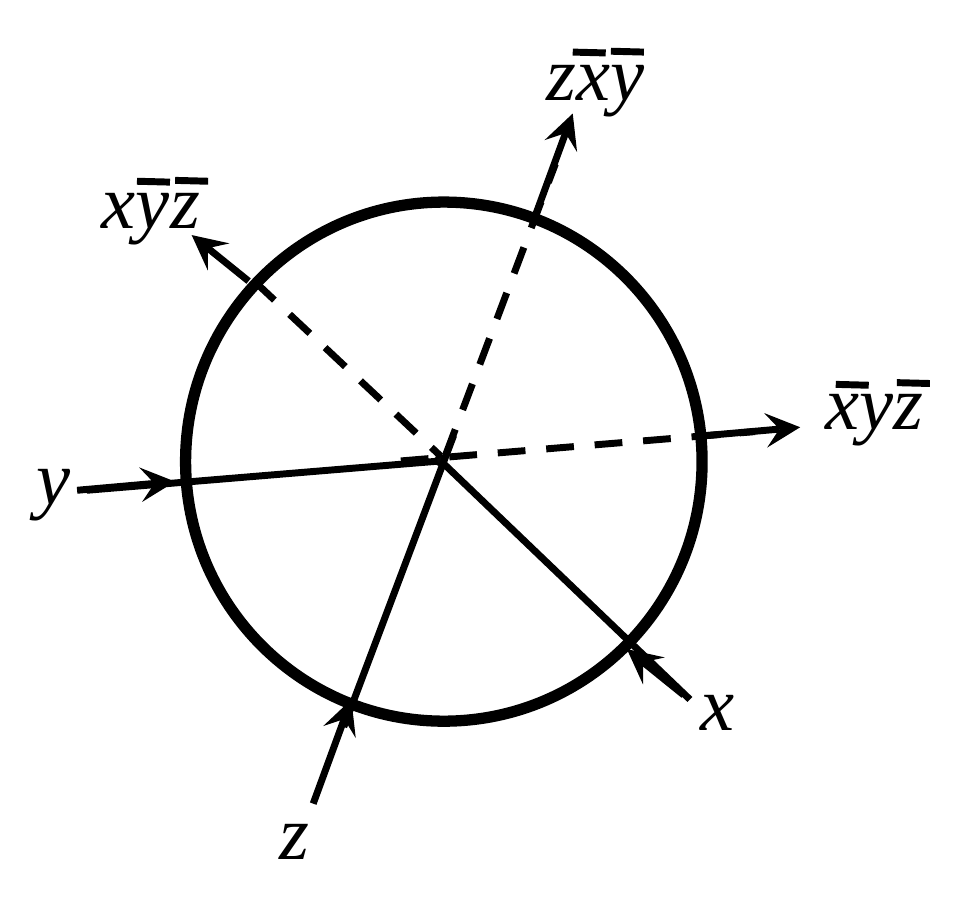}}{\text{\footnotesize high }\phi}$
}
\caption{Two types of logical gates controllable by illumination level $\phi$.}
\label{gates}
\end{figure}

We consider two types of gates implemented in BZ-vesicles via the collision of wave-fragments (Fig.~\ref{gates}).

\vspace{0.5cm}

\begin{proposition}
Let Boolean values of $x$ and $y$ be represented by wave-fragments then a BZ-vesicle implements a two-input three-output
switchable logical gate $\langle x, y, \phi  \rangle \rightarrow \langle x \overline{y}, \chi(\phi) xy,  \overline{x}y \rangle$
where $\chi(\phi)=1$ ({\sc True}) if $\phi = \phi_{\text low}$, and 0 ({\sc False}) otherwise.
\end{proposition}

\begin{figure}[!tbp]
\centering
\subfigure[$\alpha=\frac{7\pi}{18}, \phi=0.07871$]{
$\stackrel{\includegraphics[width=0.23\textwidth]{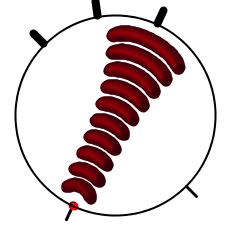}}{\scriptstyle x=1, \, y=0}$
$\stackrel{\includegraphics[width=0.23\textwidth]{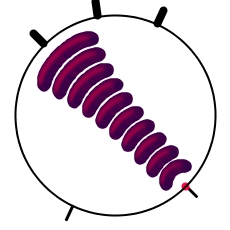}}{\scriptstyle x=0,\,  y=1}$
$\stackrel{\includegraphics[width=0.23\textwidth]{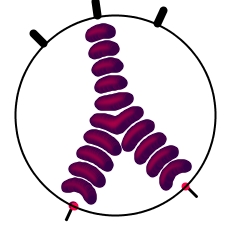}}{\scriptstyle x=1,\,  y=1}$
}
\subfigure[$\alpha=\frac{7\pi}{18}, \phi=0.07873$]{
$\stackrel{\includegraphics[width=0.23\textwidth]{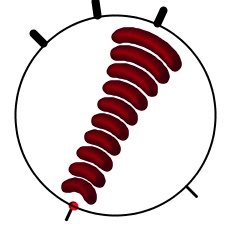}}{\scriptstyle x=1,\,  y=0}$
$\stackrel{\includegraphics[width=0.23\textwidth]{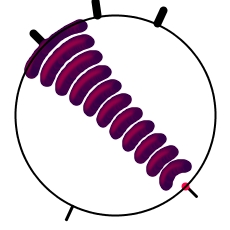}}{\scriptstyle x=0,\,  y=1}$
$\stackrel{\includegraphics[width=0.23\textwidth]{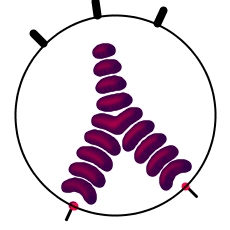}}{\scriptstyle x=1,\,  y=1}$
}
\subfigure[$\alpha=\frac{2\pi}{3}, \phi=0.07874$]{
$\stackrel{\includegraphics[width=0.23\textwidth]{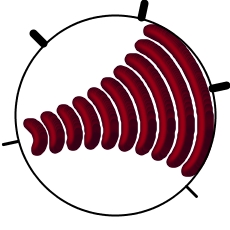}}{\scriptstyle x=1,\,  y=0}$
$\stackrel{\includegraphics[width=0.23\textwidth]{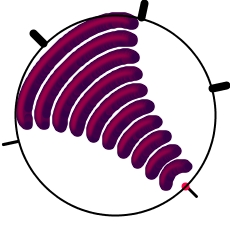}}{\scriptstyle x=0,\,  y=1}$
$\stackrel{\includegraphics[width=0.23\textwidth]{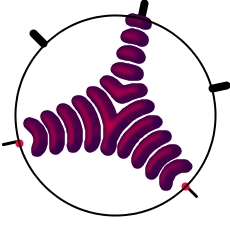}}{\scriptstyle x=1,\,  y=1}$
}
\subfigure[$\alpha=\frac{2\pi}{3}, \phi=0.07875$]{
$\stackrel{\includegraphics[width=0.23\textwidth]{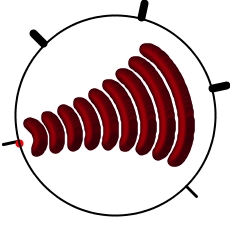}}{\scriptstyle x=1,\,  y=0}$
$\stackrel{\includegraphics[width=0.23\textwidth]{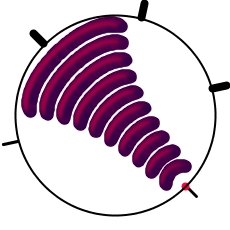}}{\scriptstyle x=0,\,  y=1}$
$\stackrel{\includegraphics[width=0.23\textwidth]{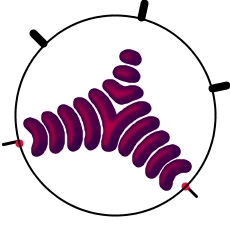}}{\scriptstyle x=1,\,  y=1}$
}
\caption{Implementation of polymorphic gate $\langle x, y, \phi  \rangle \rightarrow \langle x \overline{y}, \chi(\phi) xy,  \overline{x}y \rangle$. Scheme of the gate is shown in Fig.~\ref{gates}a. Time lapse snapshots of the sub-excitable media are shown for various illumination levels $\phi$ and collision angles $\alpha$.  Wave-fragments represent logical values of inputs $x$ and $y$.  Inputs (entry points, pores) are marked by thin lines, outputs (exit points, pores) are marked by thick lines.}
\label{gateexample}
\end{figure}

Let there be a maximum of two wave-fragments entering a BZ-vesicle. The wave-fragments 
enter the vesicle along trajectories $x$ and $y$ (Fig.~\ref{gates}a). We assume that presence of a wave-fragment 
at entry point $x$ represents a logical value {\sc True}, absence --- logical value {\sc False}. Similarly, 
if there is a wave-fragment entering BZ-vesicle along trajectory $y$ we assume $y$={\sc True}, otherwise $y$={\sc False}.
When just one of the input values is {\sc True} then the solitary wave-fragment passes through the vesicle without significant modification and
exits the vesicle at the site opposite its entry point (Fig.~\ref{gateexample}a--d, $x=1, y=0$ and $x=0, y=1$). If two wave-fragments enter the vesicle they interact and do not follow their original trajectories. Thus the output trajectories along which the undisturbed wave-fragments $x$ and $y$ move represent functions $x \overline{y}$ and $\overline{x} y$, respectively (Fig.~\ref{gates}a).

Interaction of wave-fragments is determined by level of illumination. When illumination is low enough, 
say $\phi_{\text low}$, the colliding wave-fragments merge in a new, i.e. travelling along new trajectory, 
wave-fragment (Fig.~\ref{gateexample}a and c, $x=1, y=1$). The new wave-fragment exiting the BZ-vesicle represents
operation $xy$ (Fig.~\ref{gates}a, left). For higher level of illumination, say $\phi_{\text high}$, the colliding 
wave-fragments annihilate each other (Fig.~\ref{gateexample}b and d, $x=1, y=1$), no additional operation is realised.

This is true for wave-fragments colliding at almost any angle over $\pi/6$. However for any particular angle we must 
select unique values of $\phi_{\text{low}}$ and $\phi_{\text{high}}$.

\begin{figure}[!tbp]
\centering
\subfigure[$\phi=0.0784$]{
$\stackrel{\includegraphics[width=0.2\textwidth]{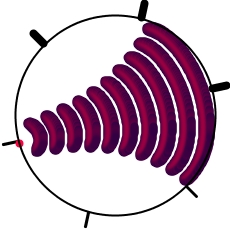}}{\scriptstyle x=1,y=0,z=0}$
$\stackrel{\includegraphics[width=0.2\textwidth]{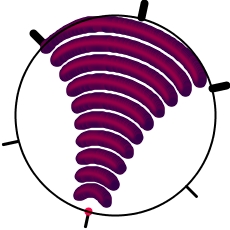}}{\scriptstyle x=0,y=0,z=1}$
$\stackrel{\includegraphics[width=0.2\textwidth]{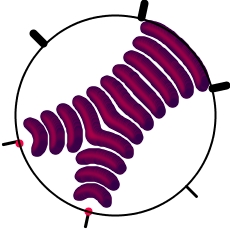}}{\scriptstyle x=1,y=0,z=1}$
$\stackrel{\includegraphics[width=0.2\textwidth]{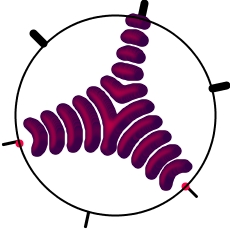}}{\scriptstyle x=1,y=0,z=1}$
$\stackrel{\includegraphics[width=0.2\textwidth]{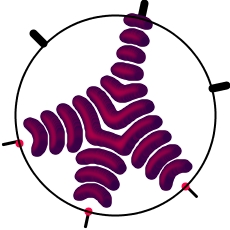}}{\scriptstyle x=1,y=1,z=1}$
}\\
\subfigure[$\phi=0.07849$]{
$\stackrel{\includegraphics[width=0.2\textwidth]{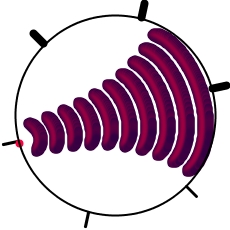}}{\scriptstyle x=1,y=0,z=0}$
$\stackrel{\includegraphics[width=0.2\textwidth]{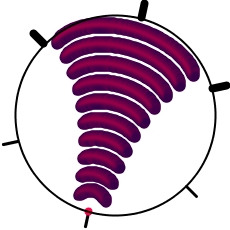}}{\scriptstyle x=0,y=0,z=1}$
$\stackrel{\includegraphics[width=0.2\textwidth]{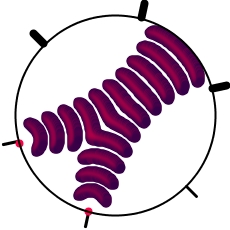}}{\scriptstyle x=1,y=0,z=1}$
$\stackrel{\includegraphics[width=0.2\textwidth]{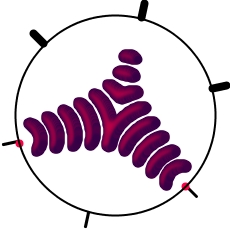}}{\scriptstyle x=1,y=1,z=0}$
$\stackrel{\includegraphics[width=0.2\textwidth]{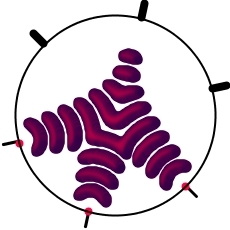}}{\scriptstyle x=1,y=1,z=1}$
}\\
%\subfigure[$\phi=0.07849$]{
%$\stackrel{\includegraphics[width=0.21\textwidth]{figs/xyz-gate/high_x00}{\scriptstyle x=1,y=0,z=0}$
%$\stackrel{\includegraphics[width=0.21\textwidth]{figs/xyz-gate/high_0z0}{\scriptstyle x=0,y=1,z=0}$
%$\stackrel{\includegraphics[width=0.21\textwidth]{figs/xyz-gate/high_xz0}{\scriptstyle x=1,y=1,z=0}$
%$\stackrel{\includegraphics[width=0.21\textwidth]{figs/xyz-gate/high_x0y}{\scriptstyle x=1,y=0,z=1}$
%$\stackrel{\includegraphics[width=0.21\textwidth]{figs/xyz-gate/high_xzy}{\scriptstyle x=1,y=1,z=1}$
%}
\caption{Implementation of polymorphic gate $\langle x,  z, y, \phi  \rangle \rightarrow 
\langle x \overline{y} \, \overline{z}, \chi(\phi) z(\overline{x \oplus y}) + \overline{\chi(\phi)}\overline{x} \,\overline{y} z,  \overline{x} y \overline{z} \rangle$. Scheme of the gate is shown in Fig.~\ref{gates}b. Time lapse snapshots of sub-excitable medium are shown for various illumination levels $\phi$ and collision angles $\alpha$.  Waves represent logical values of inputs $x$ and $y$.  Wave-fragments $x$ and $z$, and $z$ and $y$ collide at angle $\frac{\pi}{6}$; wave-fragments $x$ and $y$ collide at angle $\frac{\pi}{3}$. Inputs (entry points, pores) are marked by thin lines, outputs (exit points, pores) are marked by thick lines.}
\label{threewaves}
\end{figure}

\vspace{0.5cm}

\begin{proposition}
Let Boolean values of $x$, $y$ and $z$ be represented by wave-fragments then a BZ-vesicle implements a three-input three-output
switchable logical gate $\langle x,  z, y, \phi  \rangle \rightarrow 
\langle x \overline{y} \, \overline{z}, \chi(\phi) z(\overline{x \oplus y}) + \overline{\chi(\phi)}\overline{x} \,\overline{y} z,  \overline{x} y \overline{z} \rangle$
where $\chi(\phi)=1$ ({\sc True}) if $\phi = \phi_{\text low}$, and 0 ({\sc False}) otherwise.
\end{proposition}

Outputs presented by trajectories of undisturbed signals $x$ ($y$) --- $x \overline{y} \, \overline{z}$ 
($\overline{x} y \overline{z}$) --- are determined as follows. Wave-fragment $x$ ($y$) continues traveling 
along its original trajectory only if neither wave-fragment $y$ ($x$) nor wave-fragment $z$ enter the 
vesicle (Fig.~\ref{gates} and Fig.~\ref{threewaves}a and b, $x=1$, $y=0$, $z=0$ and $x=0$, $y=1$, $z=0$). 

The following scenarios take place for both low $\phi_{\text low}$ and high $\phi_{\text high}$ levels of 
illumination. If only wave-fragment $z$ is present, it travels through the vesicle undisturbed (Fig.~\ref{threewaves}, 
$x=0$, $y=0$, $z=1$). When wave-fragment $z$ is present and also either wave-fragment $x$ or $y$ the wave-fragments 
collide and form a wave-fragment whose velocity vector is an average of the velocity vectors of the colliding 
wave-fragments. \emph{The newly formed wave-fragment collides with the vesicle's wall just between the output channels 
and misses both of the potential exit points.} Thus no output is generated (Fig.~\ref{threewaves},  $x=1, y=0, z=1$ and $x=0, y=1, z=1$).

For two combinations of inputs ---  $x=1, y=1, z=0$ and $x=1, y=1, z=1$ --- the outcomes depend on the level of illumination.
If only wave-fragments $x$ and $y$ or all three wave-fragments enter the vesicle they collide and annihilate when the 
level of illumination is high $\phi_{\text high}$ (Fig.~\ref{threewaves}a). The fragments merge and form a new 
wave-fragment, which hits the output channel thus generating output value {\sc True}, when the level of illumination is low $\phi_{\text high}$ (Fig.~\ref{threewaves}b). Thus the output channel opposite to the input channel $z$ generates 
$z(\overline{x \oplus y})$ when the level of illumination is low, and it generates  $\overline{x} \,\overline{y} z$
when the level of illumination is high.

By assigning constant {\sc True} to input $z$, we realize a two-input one-output gate  
$\langle x, y, \phi  \rangle \rightarrow 
\langle \chi(\phi) z(\overline{x \oplus y}) + \overline{\chi(\phi)}\overline{x} \,\overline{y} z \rangle$. 
Thus we arrive at the main finding of the current paper:

\vspace{0.5cm}

\begin{proposition}
Fragments travelling and colliding within a BZ-vesicle implement a polymorphic logical gate switchable between functional states {\sc xnor} and {\sc nor} by changing the degree of illumination.  
\end{proposition}

\section{Conclusion}

In a numerical model of the light-sensitive Belousov-Zhabotinsky (BZ) medium in a sub-excitable mode localized traveling excitation waves 
are formed. We interpreted these localizations as quanta of information, values of logical variables. When two or more localizations
collide they annihilate or form a new localization. We interpreted post-collision trajectories of the localizations
as the results of a computation. We demonstrated that by colliding wave-fragments in an encapsulated excitable chemical medium we
can realise a number of logical gates. We showed that by changing the illumination of the chemical medium we could switch between 
different outcomes of the computation. Thus we were able to  realise a polymophic logical gate which could execute either function {\sc xnor} or {\sc nor} depending on the level of illumination. Gate {\sc nor} is a universal gate, thus, as a byproduct, we demonstrated the computational universality of the BZ medium when in a subexcitable state.

We would like to outline two main directions of further studies. In the theoretical part, we aim to focus on
cascading BZ-based polymorphic gates into larger logical circuits and arithmetic schemes. In experimental part, 
we aspire to implement theoretical constructs in chemical laboratory experiments.

\section{Acknowledgements}

The work is part of the European project 248992 funded under 7th FWP (Seventh Framework Programme) FET Proactive 3: Bio-Chemistry-Based Information Technology CHEM-IT (ICT-2009.8.3). We thank the project coordinator Peter Dittrich and project partners Jerzy Gorecki and Klaus-Peter Zauner for their inspirations and useful discussions.


\clearpage

\begin{thebibliography}{99}

\bibitem[Adamatzky, 2003]
{cbc}
Adamatzky~A. (Ed.) Collision-Based Computing. 
Springer, 2003.


	\bibitem[Adamatzky, 2004]
	{adamatzky_2004_collision} Adamatzky A. Collision-based
  computing in Belousov--Zhabotinsky medium. Chaos Solitons Fractals
  21 (2004) 1259--1264.

	\bibitem[Adamatzky \& De Lacy Costello, 2007]
	{andy_ben_BZ_collision} Adamatzky A., and De Lacy Costello B.
  Binary collisions between wave-fragments in a sub-excitable
  Belousov-Zhabotinsky medium. Chaos, Solitons \& Fractals 34 (2007)
  307--315.
  
	\bibitem[Adamatzky, 2010]
	{adamatzky_physarumgates} Adamatzky~A.  Slime mould logical
  gates: exploring ballistic approach (2010)  arXiv:1005.2301v1 [nlin.PS]
  (2010).  \url{http://arxiv.org/abs/1005.2301}

	\bibitem[Adamatzky et al, 2010]
	{adamatzky_adder}
	Adamatzky~A., Holley~J., Bull~L., De~Lacy~Costello~B.
	On computing in fine-grained compartmentalised Belousov-Zhabotinsky medium (2010)
	arXiv:1006.1900v1  [nlin.PS] \url{http://arxiv.org/abs/1006.1900}


	\bibitem[Beato \& Engel, 2003]
	{beato_engel} Beato~V., Engel~H. Pulse propagation in a model
  for the photosensitive Belousov-Zhabotinsky reaction with external
  noise. In: Noise in Complex Systems and Stochastic Dynamics, Edited
  by Schimansky-Geier~L., Abbott~D., Neiman~A.,
  Van~den~Broeck~C. Proc. SPIE 5114 (2003) 353--362.
  
  
  
  \bibitem[De Lacy Costello et al, 2009]
  {ben_gun} De~Lacy~Costello~B., Toth~R., Stone~C.,
  Adamatzky~A., Bull~L.  Implementation of glider guns in the
  light-sensitive Belousov-Zhabotinsky medium Phys. Rev. E 79 (2009)
  026114.
  
  
  \bibitem[Berlekam et al, 1992]
  {berlekamp_1992}
	Berlekamp E.R., Conway J.H., Guy R.L.
	Winning ways for your mathematical plays, vol. 2. Academic Press; 1982.


	\bibitem[Field \& Noyes, 1974]
	{field_noyes_1974} Field~R.~J., Noyes~R.~M.  Oscillations in
  chemical systems. IV. Limit cycle behavior in a model of a real
  chemical reaction.  J. Chem. Phys. 1974 (60) 1877--1884.
  
  \bibitem[Fredkin \& Toffoli, 1982]
  {fredkin_toffoli_1982}
  Fredkin F, Toffoli T. Conservative logic. 
  Int J Theor Phys 21 (1982) 219-–253.
  
  
  \bibitem[G\'{o}recka \& G\'{o}recki, 2003]
  {gorecka_2003} G\'{o}recka J. N., G\'{o}recki J.  T-shaped
  coincidence detector as a band filter of chemical signal frequency,
  Phys. Rev. E 67 (2003) 067203.

\bibitem[G\'{o}recki et al, 2003]
{gorecki_2003} G\'{o}recki J., Yoshikawa K. and Igarashi Y.,
  On chemical reactors that can count, J. Phys. Chem. A 107 (2003)
  1664--1669.

\bibitem[G\'{o}recki et al, 2005]
{gorecki_2005} G\'{o}recki J., G\'{o}recka J. N., Yoshikawa
  K., Igarashi Y., Nagahara H.  Sensing the distance to a source of
  periodic oscillations in a nonlinear chemical medium with the output
  information coded in frequency of excitation pulses. Phys. Rev. E 72
  (2005) 046201.

\bibitem[G\'{o}recki \& G\'{o}recka, 2006]
{gorecki_2006} G\'{o}recki J. and G\'{o}recka J. N.,
  Multi-argument logical operations performed with excitable chemical
  medium, J. Chem. Phys. 124 (2006) 084101.

\bibitem[G\'{o}recki \& G\'{o}recki, 2006]
{gorecki_2006a} G\'{o}recki J., G\'{o}recka J. N.  Information
  processing with chemical excitations --- from instant machines to an
  artificial chemical brain Int J Unconv Comput 2 (2006) 321--336.

\bibitem[G\'{o}recki et al, 2009]
{gorecki_2009} G\'{o}recki~J., G\'{o}recka~J.~N., Igarashi~Y.
  Information processing with structured excitable medium, Natural
  Computing 8 (2009) 473--492.

\bibitem[G\'{o}recka et al, 2007]
{gorecka_2007} G\'{o}recka J. N., G\'{o}recki J., Igarashi Y.
  On the simplest chemical signal diodes constructed with an excitable
  medium, Int J Unconventional Computing 5 (2009) 129--143.

\bibitem[G\'{o}recki, 2010]
{gorecki_private} Gorecki~J. Private communication (2010).

\bibitem[Hjelmfelt \& Ross, 1993]
{hjelmfelt_ross_1993} Hjelmfelt A. and Ross J.
Mass-coupled chemical systems with computational properties. J.
Phys. Chem. 97 (1993) 7988--7992.

\bibitem[Hjelmfelt \& Ross, 1995]
{hjelmfelt_ross_1995} Hjelmfelt A. and Ross J.
Implementation of logic functions and computations by chemical
kinetics. Physica D 84 (1995) 180--193.

\bibitem[Hjelmfelt et al, 1993]
{hjelmfelt_1993} Hjelmfelt A., Schneider F. W. and Ross
J. Pattern recognition in coupled chemical kinetic systems.
Science 260 (1993) 335--337.

\bibitem[Hjelmfelt et al, 1991]
{hjelmfelt_1991} Hjelmfelt A., Weinberger E. D. and Ross
J. Chemical implementation of neural networks and Turing machines.
Proc. Natl. Acad. Sci. USA 88 (1991) 10983--10987.

\bibitem[Hjelmfelt et al, 1992]
{hjelmfelt_1992} Hjelmfelt A., Weinberger E. D. and Ross
J. Chemical implementation of finite-state machines. Proc. Natl.
Acad. Sci. USA 89 (1992) 383--387.


\bibitem[Lebender \& Schneider, 1994]
{lebender_1994} Lebender D. and Schneider F. W. Logical
gates using a nonlinear chemical reaction. J. Phys. Chem. 98
(1994) 7533--7537.

\bibitem[Margolus, 1984]
{margolus}
Margolus N. Physics-like models of computation. Physica D 10 (1984) 81–-95.
  
  
  \bibitem[Motoike \& Yoshikawa, 2003]
  {motoike_2003} Motoike~I.~N. and Yoshikawa~K.  Information
  operations with multiple pulses on an excitable field.  Chaos,
  Solitons \& Fractals 17 (2003) 455--461.
  
  
  \bibitem[NeuNeu, 2010]
  {neuneu} NeuNeu: Artificial Wet Neuronal Networks from
  Compartmentalised Excitable Chemical Media.  (2010)
  \url{http://neu-n.eu/}
  
  
  \bibitem[Sendi\H{n}a-Nadal et al, 2001]
  {sendina_2001} 
Sendi\H{n}a-Nadal~I., Mihaliuk~E.,
Wang~J., P\'{e}rez-Mu\H{n}uzuri~V. and Showalter~K. Wave
propagation in subexcitable media with periodically modulated
excitability. Phys. Rev. Lett. 86 (2001) 1646--1649.



\bibitem[Sielewiesiuk \& G\'{o}recki, 2001]
{sielewiesiuk_2001} Sielewiesiuk J. and G\'{o}recki J.,
  Logical functions of a cross junction of excitable chemical media,
  J. Phys. Chem., A105 (2001) 8189.
  
  \bibitem[Stoika et al, 2002]
  {stoika_2002}
  Stoica~A., Zebulum~R., Keymeulen~D. and Lohn~J.
  On polymorphic circuits and their design using evolutionary algorithms 
  


\bibitem[T\'oth et al, 1994]
{toth1994} T\'oth A., G\'asp\'ar V. and Showalter K.
Propagation of chemical waves through capillary tubes. J. Phys.
Chem. 98 (1994) 522--531.
  
  
\bibitem[T\'oth \& Showalter, 1995]
{toth_showalter} T\'{o}th A. and Showalter K. Logic gates
in excitable media. J. Chem. Phys. 103 (1995) 2058--2066.

\bibitem[Toth et al, 2009]
{RITABEN} Toth~R., Stone~C., Adamatzky~A., de Lacy
  Costello~B., Bull~L.  Experimental validation of binary collisions
  between wave-fragments in the photosensitive Belousov-Zhabotinsky
  reaction.  Chaos, Solitons \& Fractals 41 (2009) 1605--1615.
  
  
  \bibitem[Toth et al, 2009]
  {toth_2009}  
  Toth~R., Stone~C., De Lacy Costello~B., Adamatzky~A., Bull~L. 
  Simple collision-based chemical logic gates with adaptive computing. 
  J Nanotech and Molecular Computation 1 (2009) 1-13.

  
\bibitem[Yamaguchi et al, 2009]
{yoshikawa_2009} Yoshikawa~K., Motoike~I.~M., Ichino~T.,
  T. Yamaguchi, Y. Igarashi, J. Gorecki and J. N. Gorecka Basic
  information processing operations with pulses of excitation in a
  reaction-diffusion system.  Int J Unconventional Computing 5 (2009)
  3--37.
 
\bibitem[Yoshikawa et al, 2009]
{yoshikawa_2009a} Yoshikawa~K., Nagahara~H., Ichino~T.,
  J. Gorecki, J. N. Gorecka and Y. Igarashi On chemical methods of
  direction and distance sensing.  Int J Unconventional Computing 5
  (2009) 53--65.


\end{thebibliography}
\end{document}